\documentclass{article}
\usepackage{graphics}
\usepackage{epsfig}
\usepackage{amsmath}
\usepackage{graphicx}
\newcommand{\bra}[1]{\bigl\langle #1 \bigr|}
\newcommand{\ket}[1]{\bigl| #1 \bigr\rangle}
\begin{document}




\begin{center}
{\Large Communication  via entangled coherent quantum  network}
\\
A. El Allati $^{a}$, Y. Hassouni $^{a}$ N. Metwally$^{b}$\vspace{6pt} \\
$^{a}${Facult\'e des Sciences
D\'epartement de Physique, Laboratoire de Physique Th\'{e}orique
URAC 13, Universit\'e Mohammed V - Agdal. Av. Ibn Battouta, B.P.
1014, Rabat, Morocco}
\\
$^{b}${Mathematics Department,
College of Science, University of Bahrain, P.O. Box, 32038
Bahrain}
\end{center}

\begin{abstract}

A quantum network is  constructed via maximum entangled coherent
states. The possibility of using this network to achieve
communication between multi-participants is investigated. We
showed that the probability of teleported unknown state
successfully, depends on the size the used network. As the numbers
of participants increases, the successful probability does not
depend on the intensity of the field. The problem of implementing
quantum teleportation protocol via a noise quantum network is
discussed. We show one can send information perfectly with small
values of the field intensity and larger values of the noise
strength. The successful probability of this  suggested protocol
increases abruptly for larger values of the noise strength and
gradually for small values. We show that for small  size of the
used quantum network, the fidelity of the teleported state
decreases smoothly, while  it decreases abruptly for larger size
of  network.

\bigskip

\end{abstract}

\vspace{0.5cm}

\section{Introduction}

Quantum teleportation is a way to send unknown quantum state
remotely between two users, where it is destroyed at the sending
station and appears perfectly at the receiving station. Since the
first quantum teleportation protocol was first introduced by
Bennett et al \cite{CHBennet}, there are a lot of attentions have
been given to develop this phenomenon in many different
directions. Among these directions using quantum channels
described by continuous variables \cite{Vaidmann, Braunstein}.
Some efforts have been done to extent the  used quantum channel
multiparities systems \cite{karlsso, rigolin, chang}. In these
types of quantum teleportation protocols, one needs a larger
numbers of participants   collaborate together to transfer the
unknown information between any  two of them. This process is
called
  information transfer  remotely over a quantum network \cite{Lewen}.

There are some efforts have been done to generate entangled
quantum network. As an example, Nguyen  \cite{nguyen} has
constructed a quantum network consists of $2^N$ parties of
coherent states and used it to implement  quantum teleportation.
Brougham et. al \cite{Brou} have used a passive quantum networks
with logical bus topology to transfer information safely.  Also,
Ciccarello et. al \cite{CIC} have proposed a physical model for
the systematic generation of N-partite states for quantum
networking. On the other hand the presence of noise is associated
with the deterioration of performance for quantum information
tasks. Therefore investigating the dynamics of quantum network and
its usefulness to achieve quantum information protocols is very
important \cite{Plenio, Per}.

\begin{figure}
\begin{center}
  \includegraphics[scale=0.5]{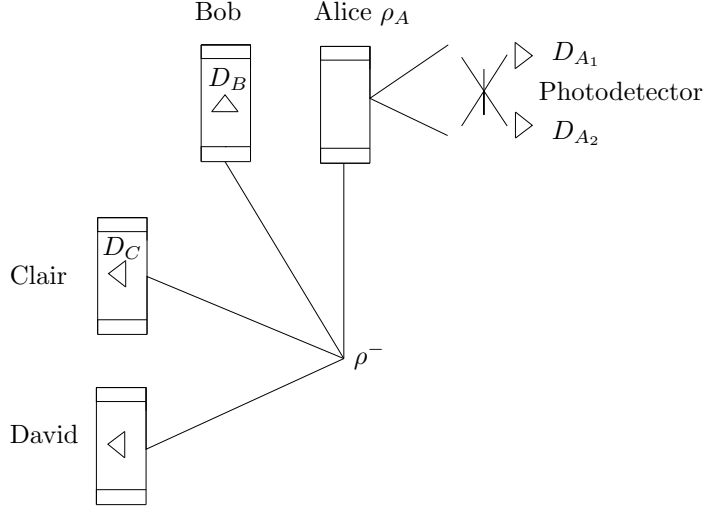}
  \put(-145,190){Alice}\put(-190,190){Bob}\put(-260,90){Clair}\put(-260,30){David}
  \put(-60,160){Photodetector}\put(-130,60){$\rho^{-}$}\put(-120,190){$\rho_{A}$}
  \put(-55,175){$D_{A_{1}}$}\put(-55,145){$D_{A_{2}}$}\put(-185,165){$D_{B}$}\put(-225,101){$D_{C}$}
  \caption{The scheme for teleporting the superposition
   coherent state $\rho_{A}$, from Alice to David within a network with assistance of other participants via quantum channel $\rho^{-}$,
   each one holds a photodetector}\label{network}
\end{center}
\end{figure}

This leads us to the aim of the current work, where we construct a
maximum  entangled coherent network, i.e. all the participants
share maximum entangled coherent states. The properties of theses
entangled coherent states have been investigated extensively in
\cite{allati}. Figure (\ref{network}) describes a quantum network
consists of four members, Alice, Bob, Clair and Diavid,   to send
unknown information between  Alice and Divid by the assistance of
Bob and Clair. This Network is generalized to includes $m$
participants and we investigate the possibility of using this
network to transfer unknown information between any two members by
the co-operation of the others. We have treated the problem of
remotely transfer information over noise quantum network, where it
is  assumed that during the construction of the network the
travelling states from the source to  their locations in the
network subject to  noise. This type of noise  is equivalent to
employing a half mirror for the noise channel \cite{Hirota}.

This article is organized as follows. In Sec.2, a  teleportation
protocol is proposed to transfer unknown information between two
distinct partners via quantum network  consists of four members.
The generalized of this protocol is discussed in Sec.2.2.
Achieving quantum teleportaion over a noise quantum network is the
subject of Sec.3. Finally, Sec.4 is devoted to discuss our
results.

\section{Telaportation via perfect quantum network}
 Entangled coherent states  have been found to be useful for
quantum teleportation. They are defined as
\begin{equation}
|\alpha\rangle_{cat}^{\pm}=N_{\alpha}^{\pm}(|\alpha\rangle\pm|-\alpha\rangle),
\end{equation}
where, $N_{\alpha}^{\pm}=[2\pm2e^{-2|\alpha|^{2}}]^{-\frac{1}{2}}$
is  a normalization factor and  $|\pm\alpha\rangle$ are two
coherent states with an equal complex amplitude $\alpha$ defined
as
\begin{equation}
|\pm\alpha\rangle=exp(-\frac{|\alpha|^{2}}{2})\sum_{n=0}^{\infty}\frac{(\pm\alpha)^{n}}{\sqrt{n!}}|n\rangle.
\end{equation}
This type of entangled coherent states have been used to teleport
a Schr\"{o}dinger cat state \cite{Enk}. The tripartites entangled
coherent state of the form
\begin{equation}\label{Ent}
\ket{\phi^{\pm}}=N^{\pm}_{\alpha}(\ket{\sqrt{2}\alpha,\alpha,\alpha}\pm\ket{-\sqrt{2}\alpha,-\alpha,-\alpha}),
\end{equation}
$N_{\alpha}=[2(1+\pm e^{-8|\alpha|^2})]^{-\frac{1}{2}}$, is used
to teleport two qubits entangled coherent state \cite{Wang}.
Recently, El-Allati et al \cite{allati} have introduced a
teleportation protocol to treleport a tripartites coherent state
via entangled coherent state of four partites  defined as:
\begin{equation}\label{MES}
\ket{\psi^{\pm}}=N^{\pm}(\ket{2\alpha,\sqrt{2}\alpha,\alpha,\alpha}\pm\ket{-2\alpha,-\sqrt{2}\alpha,-\alpha,-\alpha}),
\end{equation}
where $N^{\pm}=\sqrt{2(1\pm exp[-16|\alpha|^2])}$. Also, this
protocol has been  generalized to teleport a coherent entangled
state of $m$ modes by using entangled coherent state consists of
$m+1$ modes
 \cite{allati}. This class of states behaves as maximum entangled state, MES  for
 $\rho^-=\ket{\psi^{-}}\bra{\psi^{-}}$, where the concurrence \cite{Wootter}
 $\mathcal{C}=1$ and as partially entangled state, PES for
 $\rho^{+}=\ket{\psi^{+}}\bra{\psi^{+}}$.

 Encouraged by the work of  Nguyen \cite{nguyen}, we employ this
 class of the MES and its generalized version, which is described in \cite{allati},  to teleport unknown state
 over quantum network. We perform this idea and its generalization
 on next subsections.

\subsection{Teleportation through a network of four participants}

 Assume that Alice is asked to send unknown
 state $\rho_{A_1}$ to David, where
 \begin{equation}\label{unknown}
\rho_{A_{1}}=\frac{1}{N^{2}_{A}}(|\kappa_{1}|^{2}|2\alpha\rangle\langle2\alpha|
+ \kappa_{1}\kappa_{2}^{\ast}|2\alpha\rangle\langle-2\alpha|+
\kappa_{2}\kappa_{1}^{\ast}|-2\alpha\rangle\langle2\alpha|+|\kappa_{2}|^{2}|-2\alpha\rangle\langle-2\alpha|),
\end{equation}
 $\kappa_{1,2}$ are unknown complex numbers, and
$N_{A}=\sqrt{|\kappa_{1}|^{2}+|\kappa_{2}|^{2}+2e^{-8|\alpha|^{2}}Re(\kappa_{2}^{\ast}\kappa_{1})}$.
For this aim, the users use  quantum network consists of  four
maximum entangled coherent state given by,
 \begin{equation}\label{MES-Channel}
 \ket{\psi}_{ABCD}=N^{-}\Bigl(\ket{2\alpha,\sqrt{2}\alpha,\alpha,\alpha}_{ABCD}
-\ket{-2\alpha,-\sqrt{2}\alpha,-\alpha,-\alpha}_{ABCD}\Bigr),
\end{equation}
where $N^{-}=\sqrt{2(1- exp[-16|\alpha|^2])}$. The total state of
the system is given by
$\ket{\psi_s}=\ket{\psi}_{A_1}\otimes\ket{\psi}_{ABCD}$, where
$A,B,C$ and $D$ referee to Alice, Bob, Clair and David
respectively. To implement this protocol the network's members
achieve the following steps:

\begin{enumerate}
\item Alice  mixes the unknown state,
$\rho_{A_{1}}=\ket{\psi}_{A_1}\bra{\psi}$ with the quantum channel
$\rho_{ABCD}= \ket{\psi}_{ABCD}\bra{\psi}$ by applying a series of
operations defined by  beam splitters  and phase shifters  as,
\cite{Wang,Rab}
\begin{equation}\label{Loc}
\rho_{A_{1}}\otimes\rho_{ABCD}^{-}\rightarrow
R_{A_{1}A}\rho_{A_{1}}\otimes\rho_{ABCD}^{-}R_{A_{1}A}^{\ast}=
\rho_{out},
\end{equation}
where $R_{ij}\bigl| \mu \bigr\rangle\bigl| \nu \bigr\rangle=\bigl|
\frac{ \mu+\nu}{\sqrt{2}} \bigr\rangle\bigl|
\frac{\mu-\nu}{\sqrt{2}} \bigr\rangle$ \cite{allati} and the
output state, $\rho_{out}$ is,
\begin{eqnarray}\label{out}
\rho_{out}&=&|\kappa_1|^2\bigl| \psi_1 \bigr\rangle\bigl\langle
\psi_1 |-|\kappa_1|^2\bigl| \psi_1 \rangle\bigl\langle \psi_2 |
-\kappa_1\kappa_2^{\ast}| \psi_1 \rangle\langle \psi_3 |+
\kappa_1\kappa_2^{\ast}| \psi_1 \rangle\langle \psi_4 |
\nonumber \\
&+&|\kappa_1|^2| \psi_2 \rangle\langle \psi_2 |
+|\kappa_1|^2\kappa_2^{\ast}| \psi_2 \rangle\langle \psi_2 |
-\kappa_1\kappa_2^{\ast}| \psi_2 \rangle\langle \psi_3 | +\kappa_1\kappa_2^*| \psi_2 \rangle\langle \psi_4 |  \nonumber \\
&-&\kappa_2\kappa_1^{\ast}| \psi_3\rangle\langle \psi_1 |
+\kappa_2\kappa_1^*| \psi_3 \rangle\langle \psi_2 | +|\kappa_3|^2|
\psi_3 \rangle\langle \psi_3 | +|\kappa_3|^2| \psi_3 \rangle\langle
\psi_4 |  \nonumber\\
&-&\kappa_2\kappa_1^{\ast}| \psi_4 \rangle\langle \psi_1 |
-\kappa_2\kappa_1^{\ast}| \psi_4 \rangle\langle \psi_2 |
-|\kappa_2|^2| \psi_4 \rangle\langle \psi_3 | +|\kappa_2|^2| \psi_4
\rangle\langle \psi_4 |,
\end{eqnarray}
with,
\begin{eqnarray}
| \psi_1 \bigr\rangle&=&| 2\sqrt{2}\alpha,0,\sqrt{2}
\alpha,\alpha,\alpha\rangle_{A_{1}ABCD},\quad | \psi_2 \rangle=|
0,2\sqrt{2}\alpha,-\sqrt{2}\alpha,-\alpha,-\alpha \bigr\rangle_{A_{1}ABCD}  \nonumber \\
\bigl| \psi_3 \bigr\rangle&=&\bigl| 0,-2\sqrt{2}\alpha,\sqrt{2}
\alpha,\alpha,\alpha \bigr\rangle_{A_{1}ABCD},\quad \bigl| \psi_4
\bigr\rangle=| -2 \sqrt{2}\alpha,0,-\sqrt{2}\alpha,-\alpha,-\alpha
\bigr\rangle_{A_{1}ABCD}.
\nonumber\\
\end{eqnarray}
\item Alice performs two photon number measurements on modes $A_1$
and $A$ using two detectors $D_{A_{1}}$ and $D_{A}$, (see
Fig.\ref{network}). Bob and Clair should carry out the local
number measurement of modes $B$ and $C$ by their detectors $D_{B}$
and $D_{C}$, respectively. There are two different possibilities
due to Alice's operations:

\begin{itemize}
  \item {\it Alice's measurements such that $n_{A_{0}}=0$,
$n_{A}>0$}

The participants, Alice, Bob and Clair send their measurement
outcomes to David via a public channel, where
$n_{A}+n_{B}+n_{C}=odd$. In this case, the state at David's hand
collapses into,
\begin{equation}\label{david1}
\rho'_{D}=\lambda_1|-\alpha\rangle\langle-\alpha|-\lambda_2|-\alpha\rangle\langle\alpha|
-\lambda_3|\alpha \rangle\langle-\alpha|+\lambda_4|
\alpha\rangle\langle\alpha|,
\end{equation}
where $\lambda_1=\frac{|\kappa_{1}|^2}{N_1}, \lambda_2=\frac{
\kappa_{1}\kappa_{2}^{\ast}}{N_1}(-1)^{n_{A}+n_{B}+n_{C}},
\lambda_3=\frac{
\kappa_{2}\kappa_{1}^{\ast}}{N_1}(-1)^{n_{A}+n_{B}+n_{C}},
\lambda_4=\frac{|\kappa_{2}|^2}{N_1} $, and
$N_1=|\kappa_{1}|^{2}+|\kappa_{2}|^{2}-2(-1)^{n_{A}+n_{B}+n_{C}}e^{-2|
\alpha|^{2}}Re(\kappa_{2}^{\ast}\kappa_{1})$ is the normalized
factor.

Finally, David applies the operator $P(\pi)$ (phase shifter) to
Eq.(\ref{david1}) to get the final state
$\rho_{D}=P(\pi)\rho'_{D}P^{*}(\pi)$, which is exactly the state
$\rho_0=\ket\psi_0\bra\psi$, with a  probability of success in
given by,
\begin{equation}\label{p1}
\mathcal{P}_{1} =
                \frac{e^{-3|\alpha|^{2}}}{4\sinh(8|\alpha|^{2})}\big\{\sinh(11|\alpha|^{2}-\sinh(3|\alpha|^{2})\big\}.
\end{equation}

  \item {\it Alice's measurements such that $n_{A_{1}}>0$,
  $n_{A_{2}}=0$.}

  In this case  David's state collapse into,
\begin{equation}\label{david2}
\rho''_{D}=\lambda_1|\alpha\rangle\langle\alpha|-\lambda_2|\alpha\rangle\langle-\alpha|
-\lambda_3|-\alpha \rangle\langle\alpha|+\lambda_4|
-\alpha\rangle\langle-\alpha|,
\end{equation}
where $\lambda_1=\frac{|\kappa_{1}|^2}{N_1}, \lambda_2=\frac{
\kappa_{1}\kappa_{2}^{\ast}}{N_1}(-1)^{n_{A_{1}}+n_{B}+n_{C}},
\lambda_3=\frac{
\kappa_{1}^{\ast}\kappa_{2}}{N_1}(-1)^{n_{A_{1}}+n_{B}+n_{C}},
\lambda_4=\frac{|\kappa_{2}|^2}{N_1}$ and
$N_1=|\kappa_{1}|^{2}+|\kappa_{2}|^{2}-2(-1)^{n_{A_{1}}+n_{B}+n_{C}}e^{-2|
\alpha|^{2}}Re(\kappa_{2}^{\ast}\kappa_{1})$ is the normalized
factor.

However if  $n_{A}+n_{B}+n_{C}$ is odd, then nothing should be
done by David  and the teleported state is obtained with a
probability of success,
\begin{equation}\label{p2}
\mathcal{P}_{2}
=\frac{e^{-3|\alpha|^{2}}}{4\sinh(8|\alpha|^{2})}\big\{\sinh(11|\alpha|^{2})-\sinh(3|\alpha|^{2})\big\},
\end{equation}
From Eqs.(\ref{p1}$\&$\ref{p2}), the   probability in both cases
are equal, $\mathcal{P}_{1}=\mathcal{P}_{2}$. Therefor the total
probability of successful teleportation is given by,

\begin{equation}
\mathcal{P}=\frac{e^{-3|\alpha|^{2}}}{2\sinh(8|\alpha|^{2})}\big\{\sinh(11|\alpha|^{2})-\sinh(3|\alpha|^{2})\big\},
\end{equation}

which tend to $\frac{1}{2}$ in the limit $|\alpha|\rightarrow0$
and to $\frac{1}{2}$ in the limit $|\alpha|\rightarrow\infty$.

\item At the end, David can makes some operators on the bases
$|\alpha\rangle\rightarrow|2\alpha\rangle$ and
$|-\alpha\rangle\rightarrow|-2\alpha\rangle$ using the modified
beam splitter for producing exactly the state of Alice
$\rho_{D}=\rho_{A}$.
\end{itemize}
\end{enumerate}
 Figure.(2), describes the behavior of the succusses probability,
 $\mathcal{P}$ of achieving the quantum teleportation protocol via
 a network consists of four members sharing a maximum entangled
 coherent state  defined by (\ref{out}). It is clear that, for
 small values of the mean photon number i.e. $|\alpha^2| \in[0,0.7]$ the
 probability decreases as $|\alpha^2|$ increases.
However, the minimum value of
 $\mathcal{P}$ in the interval $\alpha\in[0,0.7]$ is almost  $\simeq
 0.43$ and  in the limit $\alpha\rightarrow 0$, the probability
 tends to $0.5$.  On the other hand, the probability of successes is
 independent of $\alpha$ for $|\alpha|^2>1$, where
 $\mathcal{P}=0.5$.
Comparing our results with that  depicted in \cite{nguyen}, we can
see that constructing a quantum network by using a maximum
entangled state defined by (\ref{Ent}) is much better.  It is
clear that in the earlier study \cite{nguyen},
$\mathcal{P}\rightarrow 0.25$ in the limit $\alpha\rightarrow 0$,
while in the current investigation $\mathcal{P}\rightarrow 0.5$ as
$\alpha\rightarrow 0$. The second reason, the probability of
successful teleportation is independent from the intensity of the
field, for $|\alpha|^2<0.7$, while in \cite{nguyen} for
$|\alpha|^2<3$. Third  reason although, $\mathcal{P}$ decreases in
the interval $\alpha\in[0,0.7]$, the minimum probability still
much better than that depicted in the earlier work.

\begin{figure}
\begin{center}
  \includegraphics[width=19pc,height=12pc]{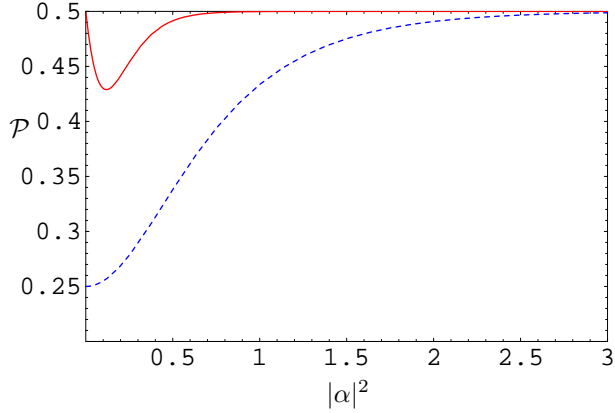}
  \put(-230,90){$\mathcal{P}$}\put(-110,-10){$|\alpha|^{2}$}
  \caption{The probability of successful teleportation $\mathcal{P}$ as function of $|\alpha|^{2}$ by continue line
  and compared it in the same case to the probability of successful of Nguyen \cite{nguyen} which defined by dash line.}\label{davidpro1}
\end{center}
\end{figure}

\subsection{Teleportation through a network of $m$ participants}
These results can be generalized to $m+1$  participants on the
quantum network by  using maximally entangled coherent states. For
this aim, we assume that the network's users share maximally
entangled coherent state of the form
\begin{equation}\label{MES}
|\Psi\rangle^{-}_{net}=
N_{m+1}(|2^{\frac{m-1}{2}}\alpha\rangle_{m}...|2^{\frac{1}{2}}\alpha\rangle_{2}|\alpha\rangle_{1}|\alpha\rangle_{0}
-|-2^{\frac{m-1}{2}}\alpha\rangle_{m}...|-2^{\frac{1}{2}}\alpha\rangle_{2}|-\alpha\rangle_{1}|-\alpha\rangle_{0}),
\end{equation}
with $N_{m+1}$ is the normalization factor  given by
\begin{equation}
N_{m+1}=[2(1- e^{-2^{m+1}|\alpha|^{2}})]^{-\frac{1}{2}}
\end{equation}
The entangled and separable properties of this class of states are
investigated in \cite{allati}. Also, these states are employed to
perform a teleportation between any number of parties. In this
context, we use them  to achieve quantum teleportation over a
network. The suggested protocol is implemented as following.

\begin{enumerate}
\item
 Assume that the user $m$,  who shares a maximum entangled state (\ref{MES}) with the remaining
 $m-1$  members, co-operate to send  the state $\ket{\psi_0}$ to any member of the network. In this
 case, the generator of this type of the MES, sends the mode $m$ to
 the emitter and the modes $m-1$ to remaining users. The total state of the system is  defined by
 $\rho_s=\rho_{A_0}.\rho_{net}$ where $\rho_{net}=\ket{\psi}_{net}\bra{\psi_{net}}$.

\item The user $m$ performs a sequence of local operations
(\ref{Loc}) on his own state and the state $\ket{\psi_0}$. Then by
using the two  detectors $D_0$ and $D_m$, the user $m$ counts the
photon numbers in modes $0$ and $m$.

\item The other users carry out the local number measurement of
modes $ m-1, m-2..3,2,1$ by the local detectors $D_{m-1}$,
$D_{m-2}$....$D_{1}$, respectively. As a result of these
measurements there are two possibilities:
 {\it The first possibility is obtained for
$n_{0}=0$, $n_{m}>0$}

 In this case, the state $\ket{\psi_0}$
collapses  at the receiver into,
\begin{equation}\label{davidg1}
\rho'_{0}=\lambda_1|-\alpha\rangle\langle-\alpha|-\lambda_2|-\alpha\rangle\langle\alpha|
-\lambda_3|\alpha \rangle\langle-\alpha|+\lambda_4|
\alpha\rangle\langle\alpha|,
\end{equation}
where $\lambda_1=\frac{|\kappa_{1}|^2}{N_{m}}, \lambda_2=\frac{
\kappa_{1}\kappa_{2}^{\ast}}{N_{m}}(-1)^{n_{m}+n_{m-1}..+n_{1}},
\lambda_3=\frac{
\kappa_{2}\kappa_{1}^{\ast}}{N_{m}}(-1)^{n_{n_{m}+n_{m-1}..+n_{1}}},
\lambda_4=\frac{|\kappa_{2}|^2}{N_{m}} $ and
$N_{m}=|\kappa_{1}|^{2}+|\kappa_{2}|^{2}-2(-1)^{n_{m}+n_{m-1}..+n_{1}}e^{-2|
\alpha|^{2}}Re(\kappa_{2}^{\ast}\kappa_{1})$ is the normalized
factor.

 The remaining  participants send  their classical  results
 via a public  channel to the receiver, where
$n_{m}+n_{m-1}..+n_{1}=odd$, who applies the operator $P(\pi)$ and
a series of operations $R_{i,j}$ defined by modified beam splitter
for replica the original state on the state (\ref{davidg1}) to get
the state $\rho_{0}=P(\pi)\rho'_{0}P^{*}(\pi)$.

The second possibility:{\it  the measures of the partners such
that, $n_{m+1}>0$, $n_{m}=0$},  in this case, one gets the same
probability of successes as the previous case. So the information
is transformed with   an average probability given by,
\begin{equation}\label{Prob2}
\mathcal{P}=\frac{e^{-(2^{m-1}-1)|\alpha|^{2}}}{2\sinh(2^{m}|\alpha|^{2})}\big\{\sinh((3*2^{m-1}-1)|\alpha|^{2})
-\sinh((2^{m-1}-1)|\alpha|^{2})\big\}.
\end{equation}

\end{enumerate}

\begin{figure}
\begin{center}
 \includegraphics[width=20pc,height=12pc]{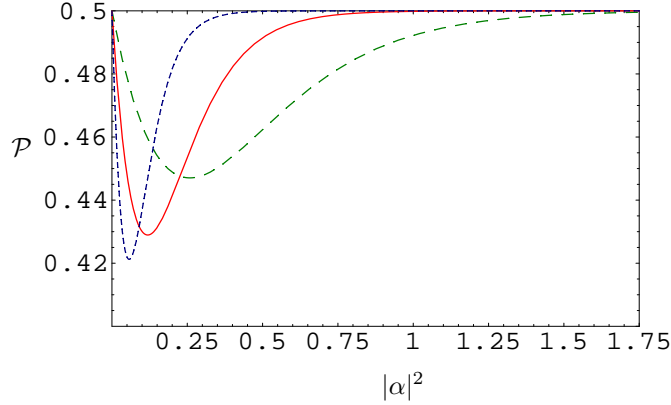}
  \put(-250,80){$\mathcal{P}$}\put(-110,-10){$|\alpha|^{2}$}
  \caption{The probability of successful teleportation $\mathcal{P}$ as function of
  $|\alpha|^{2}$, where the dash , solid and dot curves for a networks consists of $2,3,4$ participants
  respectively.}
\end{center}
\end{figure}

Fig.(3), describes  the dynamics of  the  successful probability
of teleportation (\ref{Prob2}) for different size of the quantum
network. It is clear that, for $m=2$, i.e two partners  co-operate
to send the information to the third, the probability,
$\mathcal{P}$ decreases smoothly in the interval of
$|\alpha|^2\in[0,0.25]$ and then increases gradually. However as
$|\alpha|^2$ increases the probability increases and  reaches to
its maximum value at $|\alpha|^2\simeq 1.6$. As one increases the
numbers of participants, $\mathcal{P}$ decreases abruptly in a
small range of $|\alpha|^2$  and increases faster than that
depicted  for less numbers of partners. However the minimum value
of the probability of successes increases as one decreases the
size of the network. Therefor, one can conclude that, by
increasing the number of partners in the quantum network, the
probability of success teleportation is independent of
$|\alpha|^2$. So, one can perform a quantum teleportation in the
presences of low field intensity with probability $0.5$, by
increasing the size of quantum network. On the other hand this
figure confirms that the used quantum channel (\ref{MES}) is much
better than that used in the earlier work cited in \cite{nguyen}.

\section{Teleportation via noise quantum network}
Decoherence represents the most obstacles in the context of
handling information. It arises during the interaction of the
systems with their  environments \cite{Ekert}, imperfect devices
\cite{Gero}, noise channel \cite{Sug, Metwally}, the energy loss
or the photon absorption \cite{Hirota,Enk,Hof} and etc. Therefore,
it is very important to investigate the dynamics of information in
the presence of noisy \cite{Hong}.

Investigating the properties of  quantum channels consists of
coherent states have received considerable attentions. As an
example, Van Enk \cite{Enk} has  considered the decoherence of
multi-dimensional entangled coherent states due to photon
absorption losses, where he  calculated how fast that entanglement
decays and how much entanglement is left. The  entanglement
degradation of entangled states suffering from photon absorption
losses is investigated in \cite{Enk1}. The dynamics  of multi
entangled coherent states and the possibility of using them to
perform quantum teleportation are investigated in \cite{allati}.

In this  section, we  assume that we have  quantum network, QN
consists of a multi-entangled coherent states.  Assume that we
have a source supplies the partners,  with a maximum entangled
coherent states  given by (\ref{MES}). These entangled coherent
states propagate from the source to the locations of the partners.
Due to the interaction with the environment, the maximum entangled
coherent states turn into partial entangled states, where its
degree of entanglement depends on the strength of the noise. Let
us consider that the source produces MECS defined by the density
operator, $\rho^{-}$ given by Eq.(\ref{MES}). This entangled state
turns into a partial entangled state defined by,
\begin{equation}
\rho_{PE}=U_{AE}\otimes U_{BE}\rho^{-}U^{\dagger}_{BE}\otimes
U^{\dagger}_{AE},
\end{equation}
where $U_{IE}\bigl| \alpha \bigr\rangle\bigl|
0 \bigr\rangle_E=\bigl| \sqrt{\eta}\alpha \bigr\rangle_I\bigl| \sqrt{1-\eta}%
\alpha \bigr\rangle_E, I=A$, or $B$ and $\bigl|  0 \bigr\rangle_E$
referees to the environment state. This effect is equivalent to
employing a half mirror for the noise channel \cite{Hirota}. In an
explicit form, one can write the output density operator
$\rho_{PE}$ as,
\begin{eqnarray}\label{Par}
\rho_{PE}&=& \frac{1}{{N_{\alpha}}}\Bigl[\bigl| 2\sqrt{\eta}\alpha,\sqrt{2}%
\sqrt{\eta}\alpha,\sqrt{\eta}\alpha,\sqrt{\eta}\alpha \bigr\rangle%
\bigl\langle 2\sqrt{\eta}\alpha,\sqrt{2}\sqrt{\eta}\alpha,\sqrt{\eta}\alpha,%
\sqrt{\eta}\alpha \bigr|  \nonumber \\
&+&\bigl| -2\sqrt{\eta}\alpha,-\sqrt{2}\sqrt{\eta}\alpha,-\sqrt{\eta}\alpha,-%
\sqrt{\eta}\alpha \bigr\rangle\bigl\langle -2\sqrt{\eta}\alpha,-\sqrt{2}%
\sqrt{\eta}\alpha,-\sqrt{\eta}\alpha,-\sqrt{\eta}\alpha \bigr|  \nonumber \\
&-&e^{-8(1-\eta)|\alpha|^{2}}\bigl| 2\sqrt{\eta}\alpha,\sqrt{2}\sqrt{\eta}\alpha,%
\sqrt{\eta}\alpha,\sqrt{\eta}\alpha \bigr\rangle\bigl\langle -2\sqrt{\eta}%
\alpha,-\sqrt{2}\sqrt{\eta}\alpha,-\sqrt{\eta}\alpha,-\sqrt{\eta}\alpha %
\bigr|  \nonumber \\
&-& e^{-8(1-\eta)|\alpha|^{2}}\bigl| -2\sqrt{\eta}\alpha,-\sqrt{2}\sqrt{\eta}\alpha,-%
\sqrt{\eta}\alpha,-\sqrt{\eta}\alpha \bigr\rangle\bigl\langle 2\sqrt{\eta}%
\alpha,\sqrt{2}\sqrt{\eta}\alpha,\sqrt{\eta}\alpha,\sqrt{\eta}\alpha \bigr|%
\Bigr],
\end{eqnarray}
where $N_{\alpha}=2(1-e^{-16|\alpha|^{2}})$ is the normalized
factor (for more details see \cite{allati}). Assume that the aim
of Alice  is sending unknown  state $\rho_{A_1}$ (\ref{unknown})
to Divad through the noise quantum network (\ref{Par}). The
partners follow the same  steps as described in Sec.(2.1) to
fulfill this task. If the members Alice, Bob and clair who
performed the number measurement  send their outcomes through a
classical channel to David such that, $n_{A_{1}}=0$ and
$n_{A_{2}}>0$ and $n_A+n_B+n_C$ is odd, then the final state at
Divad's hand is

\begin{figure}[b!]
\begin{center}
\includegraphics[width=20pc,height=11pc]{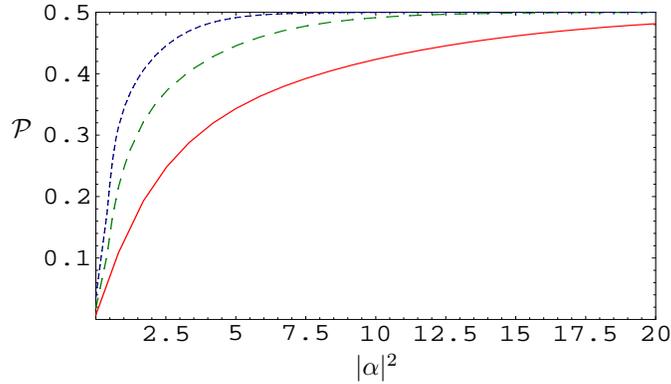}
\put(-120,-10){$|\alpha|^2$} \put(-250,80){$\mathcal{P}$}
\end{center}
\caption{The total probability, $\mathcal{P}_{t}$ of successful
quantum teleportation via noise quantum network (\ref{Par}). The
solid, dash and dot curves are evaluate for $\eta=0.02,0.05,0.1$
respectively.}
\end{figure}

\begin{eqnarray}
\rho_{D}&=&\lambda_1|-\sqrt{\eta}\alpha\rangle|-\chi\rangle\langle-\sqrt{\eta}\alpha|\langle-\chi|-\lambda_2|-\sqrt{\eta}\alpha\rangle|-\chi\rangle\langle\sqrt{\eta}\alpha|\langle\chi|  \nonumber \\
&-&\lambda_3|\sqrt{\eta}\alpha\rangle|\chi\rangle\langle-\sqrt{\eta}\alpha|\langle-\chi|+\lambda_4|
\sqrt{\eta}\alpha\rangle|\chi\rangle\langle\sqrt{\eta}\alpha|\langle\chi|,
\end{eqnarray}
where,
\begin{eqnarray}
\ket{\chi}&=&|2\sqrt{\eta^{\prime}}\alpha\rangle_{E}|\sqrt{2}\sqrt{\eta^{\prime}}
\alpha\rangle_{E}|\sqrt{\eta^{\prime}}\alpha\rangle_{E}|\sqrt{\eta^{\prime}}\alpha\rangle_{E}
\nonumber\\
\lambda_1&=&\frac{|\kappa_{1}|^2}{N_{\chi}},~ \lambda_2=\frac{
\kappa_{1}\kappa_{2}^{\ast}}{N_{\chi}}(-1)^{n_{A_{2}}+n_{B}+n_{C}},~
\lambda_3=\frac{
\kappa_{2}\kappa_{1}^{\ast}}{N_{\chi}}(-1)^{n_{A_{2}}+n_{B}+n_{C}},~
\lambda_4=\frac{|\kappa_{2}|^2}{N_{\chi}} ~
 \nonumber\\
N_{\chi}&=&|\kappa_{1}|^{2}+|\kappa_{2}|^{2}+
e^{-2|\eta\alpha|^{2}}e^{-16|\sqrt{\eta^{\prime}}\alpha|^{2}}Re(\kappa_{2}^{\ast}\kappa_{1}),
\eta^{\prime}=1-\eta.
\end{eqnarray}

The probability of finding odd number of photons in this case is
given by,
\begin{equation}
\mathcal{P}=\frac{N_{\chi}e^{-11|\sqrt{\eta}\alpha|^{2}}}{N^{2}_{A_1}
N_{-}^{2}}\big\{\sinh(11|\sqrt{\eta}\alpha|^{2})-\sinh(3|\sqrt{\eta}\alpha|^{2})\big\},
\end{equation}
where, $N_{A_1}$ is the normalization  factor of the unknown state
Eq.(5), and $N_{-}$ is the normalization of quantum channel state
(\ref{MES-Channel}).

In Fig.(4), the dynamics of the total successful probability,
$\mathcal{P}_{t}=2\mathcal{P}$ is investigated for different
values of the noise strength, $\eta$. It is clear that, as one
increases $\eta$, the probability, $\mathcal{P}_{t}$ increases
faster while  it increases gradually for small values of $\eta$.
For larger values of $|\alpha|^2$, $\mathcal{P}_t\to \frac{1}{2}$,
namely the probability of successes is independent of noise.
However for small values of the field intensity, one can increase
the probability of successes by increasing the noise strength.
Therefor to send a coded information through a noise cavity with a
large probability either one reduces the intensity of the field
and increases the noise strength or increases the field intensity
and reduces the noise strength.

\begin{figure}[b!]
\begin{center}
\includegraphics[width=15pc,height=12pc]{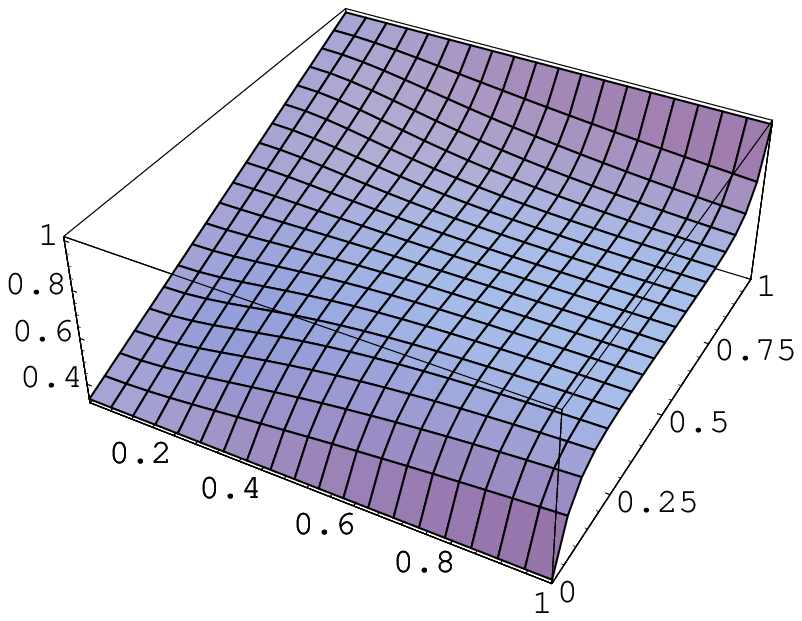}
\put(-130,15){$|\alpha|^2$}
\put(-190,70){$\mathcal{F}$}\put(-10,45){$\eta$}
 ~\quad\hspace{0.5pc}
\includegraphics[width=15pc,height=11pc]{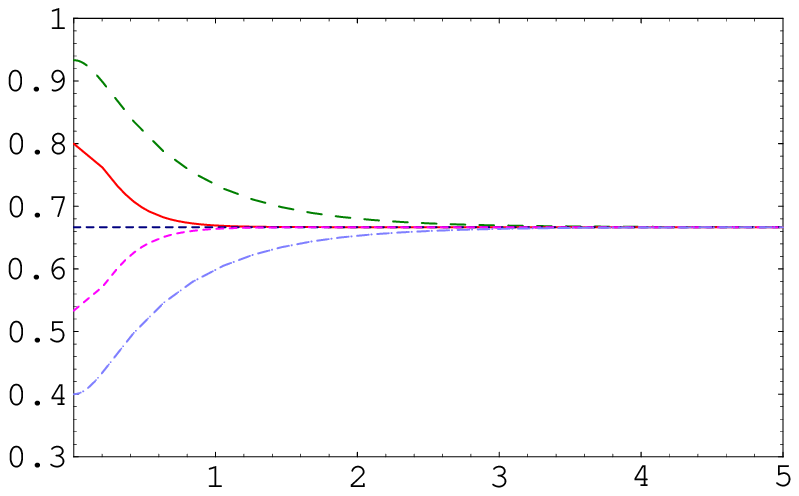}
\put(-80,-10){$|\alpha|^2$} \put(-180,65){$\mathcal{F}$}
\end{center}
\caption{(a) The fidelity of the teleported state (\ref{unknown})
by using the noise  quantum network (\ref{out})~(b) the same as
(a) but for $\eta=0.9,0.7,0.5,0.3.0.1$ from top to bottom curves.}
\end{figure}

The   fidelity, $\mathcal{F} $  of the teleported state
(\ref{unknown}),  via a  noise quantum network defined by
(\ref{Par}), is given by \cite{Horodecki,S.Oh},
\begin{equation}
\mathcal{{F}}=tr\{\tilde{\rho}_{D}\rho_{A_{1}}\}=\frac{f
N_{H}+1}{N_{H}+1},
\end{equation}
where,
\begin{equation}
f= \frac{(1-e^{-16\eta|\alpha|^{2}})(1+e^{-16(1-%
\eta)|\alpha|^{2}})}{2(1-e^{-16|\alpha|^{2}})},
\end{equation}
is the fidelity of the quantum network (\ref{out}) and $N_{H}$ is
dimension of the Hilbert space.

Fig.(5a), describes the dynamics of the fidelity, $\mathcal{F}$ in
a small interval of the field intensity, i.e.
$|\alpha|^2\in[0,1]$. It is clear that, for small values of $\eta$
and $\alpha$, the fidelity of the teleported state increases very
fast and it  increases smoothly for larger value of $\alpha$.
Therefor to send information with a unit fidelity for low
intensity field, one has to increase the noise strength. In
Fig.(5b),  the fidelity $\mathcal{F}$ is plotted  for different
values of the noise strength $\eta$.  It is shown that, for large
values of $\eta=0.95$( most upper curve), the fidelity decreases
smoothly to reaches to its lower limit
($\mathcal{F}=\frac{2}{3}$). As one decreases $\eta$, the fidelity
decreases faster and reach its lower limit at $|\alpha|^2\simeq
\frac{3}{2}$. However  for small values of $\eta\in[0,0.5]$ the
fidelity increases gradually and reaches its upper limit for
larger values of the field intensity. On the other hand, the
fidelity is independent of the noise strength in the limit of
$|\alpha|^2\rightarrow\infty$ for $\eta> 0.5$, while for any value
of $\eta\leq 0.5$, the fidelity approaches to the fidelity of
classical teleportation ,$\mathcal{F}=\frac{2}{3}$.

From the preceding discussions, one can notice that the advantage
of this current teleportation protocol is the probability of
successes is much better than that obtained by Enk et.al
\cite{Enk1}. Also, for $\eta=\frac{1}{2}$, the fidelity,
$\mathcal{F}=\frac{2}{3}$, while in the former work \cite{Enk1} (
$\mathcal{F}=\frac{1}{2})$. On the other hand, for any value of
$\alpha\in[0,1]$, the fidelity $\mathcal{F}$ is maximum  for any
value of $\eta$.

Finally, let us assume that the sender and receiver are a member
of noise  quantum network consists of $m$ participants. The
members follow the same steps as described in Sec.(2.2), they
shall end up their protocol with a final state at the receiver's
hand with a fidelity given by
\begin{equation}
\mathcal{F}_{n}=\frac{1}{3}\Bigl[\frac{(1+e^{-2^m(1-\eta)|\alpha|^2})(1-e^{-2^m\eta|\alpha|^2)}}
{(1-e^{-2^m|\alpha|^2})}+1\Bigr].
\end{equation}
In Fig.(6), we investigate the dynamics of the fidelity,
$\mathcal{F}_n$ for different size of the network, where we assume
two different values of the noise strength. In Fig.(6a), we set  a
larger value of the noise strength, $\eta=0.9$ and different size
of the quantum network. It is clear that, for small size ($m=3$),
the fidelity, $\mathcal{F}_n$ decreases  smoothly and reaches its
minimum bound for larger values of the field intensity. However
for larger values of $m$,  $\mathcal{F}_n$ decreases faster and
reaches its minimum bound for smaller values of the field
intensity. Fig.(6b), describes the dynamics of $\mathcal{F}_n$ for
a smaller value of $\eta\in[0,0.5]$, where for small values of
$m$, the fidelity increases gradually and reaches its upper bound
for larger values of $|\alpha|^2$. However for larger size of the
network., i.e $m$ is larger, the fidelity increases very fast and
reaches its upper value for small values of $|\alpha|^2$.

\begin{figure}
\begin{center}
\includegraphics[width=15pc,height=11pc]{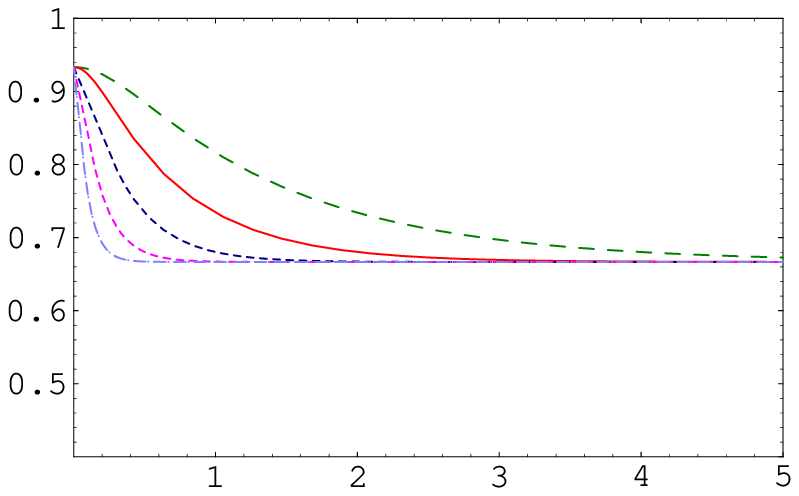}~\quad
\includegraphics[width=15pc,height=11pc]{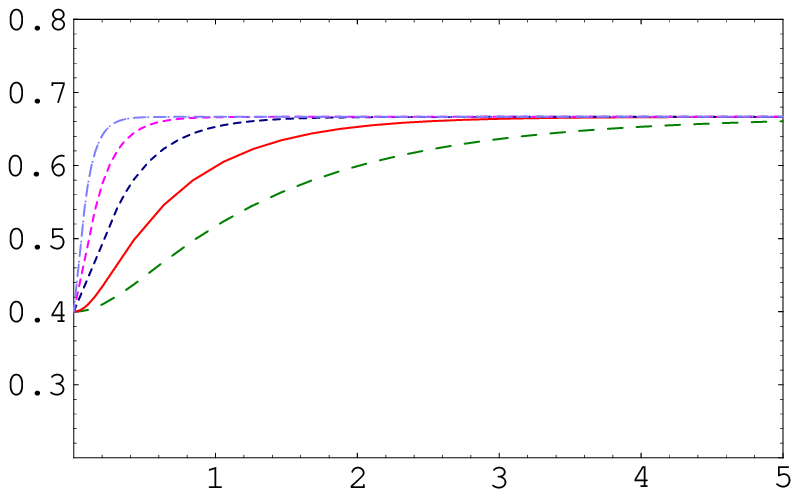}
\put(-280,-10){$|\alpha|^2$} \put(-380,65){$\mathcal{F}$}
\put(-80,-10){$|\alpha|^2$} \put(-190,65){$\mathcal{F}$}
\end{center}
\caption{(a) The fidelity of the teleported state (\ref{unknown})
by using the noise  quantum network (\ref{out})~(b) the same as
(a) but for $\eta=0.9,$ and $m=3,4,5,6,7$ from top to bottom
curves.}
\end{figure}

\section{Conclusion}

A quantum network consists of a multi-participant is constructed
by using maximum entangled coherent state. This network is used to
send information between any two members, where the other members
co-operate with the sender to achieve this aim with high
probability of successes. We showed as the size of the network
increases, i.e, the number of members are more, the probability of
the successful teleportation does not depend on the intensity of
the field. This type of entangled coherent state is much better
than that used in the earlier work of Nguyen \cite{nguyen}, where
in his proposal, the probability of successes reaches its maximum
value for larger values of the intensity of the field
($|\alpha^2|\geq 3)$, while for the current work ($|\alpha^2|\geq
0.7)$.

The possibility of using this network to perform communication
between its members in the presences of noise is investigated,
where we assume that the travelling maximum entangled state from
the sources to the locations of the members subject to noisy. Our
results  show that the probability of achieving quantum
teleportation successfully increases faster for small values of
the noise strength and reaches its maximum value for small values
of the field intensity. However for larger value of the noise
strength, the successful probability increases gradually and needs
a larger values of the field intensity to be maximum.

The effect of the noise strength and the field intensity on the
dynamics of the  fidelity of the teleported state is investigated.
We have found out that, one can send unknown information between
any two members of the network with a unit fidelity for small
values of the field intensity and larger values of the noise
strength. However the fidelity decreases as one decreases the
noise strength in $(0.5,1]$ and increases as one increases the
noise strength in$[0,0.5)$. On the other hand, the fidelity
reaches the classical limit $(\frac{2}{3})$ if we set the noise
strength $(=\frac{1}{2})$ and for larger value of the field
intensity, the fidelity of teleported state is independent of the
noise strength.

This protocol is generalized to be used  between any two members
of a network consists of $(m)$ users, where there are $(m-2)$
members should co-operate with the sender to send unknown state
safely to the receiver. Also, the possibility of implementing this
generalized protocol in the presence of noise is discussed. The
fidelity of the teleported state decreases smoothly  for small
size of the network  and abruptly for larger size.

\end{document}